\documentclass[aip,numerical,reprint,groupedaddress]{revtex4-1}

\usepackage{graphicx}
\usepackage{color}
\usepackage{amsmath}

\begin{document}

\title{Microwave-induced nonequilibrium temperature in a suspended carbon nanotube}
\author{H.L.~Hortensius}\email{h.l.hortensius@tudelft.nl}
\author{A.~\"Ozt\"{u}rk}
\author{P.~Zeng}
\author{E.F.C.~Driessen}
\author{T.M.~Klapwijk}
\affiliation{Kavli Institute of Nanoscience, Delft University of Technology, The Netherlands}

\begin{abstract}
Antenna-coupled suspended single carbon nanotubes exposed to 108 GHz microwave radiation are shown to be selectively heated with respect to their metal contacts. This leads to an increase in the conductance as well as to the development of a power-dependent DC voltage. The increased conductance stems from the temperature dependence of tunneling into a one-dimensional electron system. The DC voltage is interpreted as a thermovoltage, due to the increased temperature of the electron liquid compared to the equilibrium temperature in the leads.
\end{abstract}

\maketitle
The temperature response of carbon nanotubes exposed to microwave and terahertz radiation is of great interest both from a practical and a fundamental point of view. One of the many potential applications of carbon nanotubes lies in their response to far-infrared radiation. The bolometric response of either bundles or films of carbon nanotubes has been shown in the far- to mid- infrared range.\cite{Fu, Itkis} Recently the bolometric and rectifying response of a single carbon nanotube was studied at 77~K and at 4.2~K.\cite{SantaviccaArxiv} Also, detectors based on thermoelectric properties of suspended films of carbon nanotubes have been proposed for the terahertz range and have been demonstrated to work at optical frequencies.\cite{Benoit}

The unique one-dimensional electronic states of carbon nanotubes play an important role in the description of their DC electrical response.\cite{Deshpande} Due to the strong interactions between the electrons, the response is a collective effect of the entire electron system, as expressed in the Tomonaga-Luttinger-liquid theory.\cite{LL} In this theory, these interactions induce a reduced tunneling density of states around the Fermi energy, which leads to a power-law scaling of the conductance with bias voltage and temperature.\cite{Matveev,KaneFisher,Bockrath,Dekker} In studying transport through a carbon nanotube, an important consideration is the energy relaxation. In recent experimental work, indications of weak energy relaxation were reported.\cite{Chen} In subsequent theoretical work, the intrinsic relation between weak energy relaxation and one-dimensional physics was emphasized.\cite{Bagrets,Karzich} It is to be expected that the conductivity of a carbon nanotube will also depend on this weak energy relaxation, when exposed to radiation.

In previous work, carbon nanotubes in direct contact with a substrate were studied. This contact leads to interaction with charges in surface dielectrics, an increase in the scattering probability, and phonon-exchange with the substrate.\cite{Zhong,Glazman,Steiner} In order to avoid these potential complications and increase the thermal response, we have developed \emph{suspended} carbon nanotubes in broadband antennas. We describe the realization of these suspended carbon nanotubes and their response to 108 GHz radiation. Two main effects are observed and analyzed: the differential conductance of the nanotube is described by an increased effective temperature, and a thermoelectric offset voltage develops due to the strong gradients in the electron temperature profile at the nanotube-lead interfaces.

\begin{figure}[h!]
\includegraphics{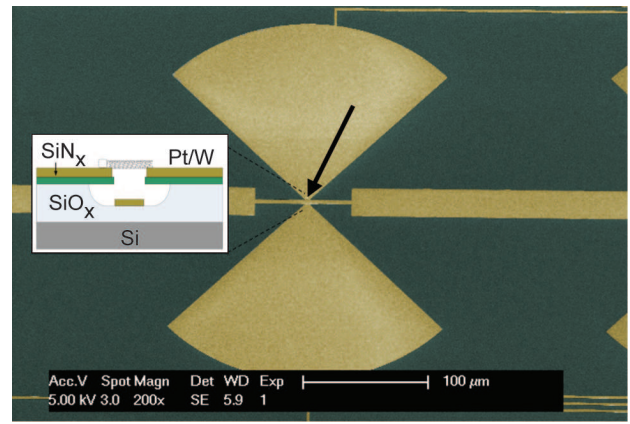}
\caption{\label{fig:sample} (Color online) Scanning electron microscope image of a suspended carbon nanotube sample. The Pt/W layer of the antenna and gate is shown in yellow (light gray), while the silicon nitride substrate is shown in green (dark gray). The iron-molybdenum catalyst particle is indicated by the arrow. The gate is located in a 250 nm deep trench. The inset shows a schematic representation of a cutthrough at the nanotube location.}
\end{figure}

Fig.~1 shows a schematic layout of the sample used. An antenna pattern is made with a catalyst particle which is used as a seed to locally grow the nanotubes. One of these can bridge the trench between the antenna pads, a methodology pioneered by Cao et al.\cite{Dai} and Steele et al.\cite{Gary}. The substrate is highly resistive silicon with a 490 nm silicon oxide layer and a 38 nm silicon nitride layer.  First, a window is opened in the silicon nitride by $\mathrm{CHF_3/O_2}$ reactive ion etching, followed by a buffered hydrofluoric acid etch to create a 250 nm deep, 2 $\mu m$ wide trench in the silicon oxide. At the bottom of the trench, a metal gate was realized, but this was not usable in the present experiment. On top of the silicon nitride, a metallic pattern of 30 nm platinum with a 10 nm tungsten adhesion layer is evaporated, which serves both as an antenna and as DC electrodes. The nanotubes are grown in the final step of fabrication by chemical vapor deposition from the iron-molybdenum catalyst seed, which is positioned at 1 $\mathrm{\mu m}$ from the trench using a lift-off technique. In one fabrication run 9 samples are made with 30 antenna pairs each. About 5\% of the antenna pairs yield a finite resistance, indicating the presence of a nanotube. The results presented here all originate from one specific sample. Similar results were obtained for five samples.

The measurement setup consists of a helium cryostat with optical access, as shown in the bottom-right inset of Fig.~2. The sample is mounted on a copper block, allowing a variation of the temperature from 5 K to 20 K by heating of a resistor. Microwave radiation is generated by a YIG-oscillator at 18 GHz and an active multiplier chain which multiplies the frequency to 108 GHz. The power is varied by changing the bias voltage of an amplifier which is part of the multiplier chain. The maximum power incident on the sample is estimated to be 83 nW/mm$^2$, based on a calibration with a pyro-electric detector placed at the position of the cryostat. The radiation passes through three high-density poly-ethylene windows at room temperature, 77 K, and 4.2 K, respectively. These windows are transmissive up to $\sim 90\%$ at the frequencies used in these experiments. The total area covered by one bowtie antenna is about 0.1 mm$^2$, leading to the estimated maximum power incident on the area of the bowtie antenna of 8.3 nW. In the remainder of the article, we will refer to this estimate as the incident microwave power $P_\mathrm{MW}$. Due to the large impedance mismatch between free space and the carbon nanotube, only a small fraction of the incident power is absorbed in the nanotube.

\begin{figure}[h!]
  \includegraphics{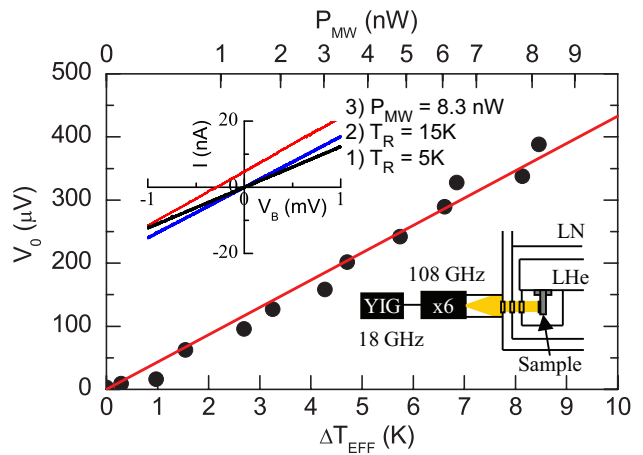}
   \caption{\label{fig:DC} (Color online) Zero-current offset voltage as a function of the temperature increase in the nanotube under microwave irradiation (bottom axis) and the incident power on the antenna (top axis). The top-left inset shows the low-bias current-voltage characteristics. Curve 1 (black) is for the suspended carbon nanotube at 5~K, curve 2 (blue) is taken under heating of the entire substrate-lead-nanotube system to 15~K, and curve 3 (red) is taken under microwave irradiation with the bath temperature 5 K. The bottom-left inset shows a schematic representation of the experimental setup with microwave radiation incident on the sample.}
\end{figure}

The top left inset of Fig.~2 shows the small-signal current-voltage characteristics of the nanotube for two different bath temperatures as well as one in the presence of radiation. Clearly, the resistances of the nanotube are constant over this bias voltage range. Curve 1 (black) is measured at a bath temperature of $5$~K, without  applied radiation. Curve 2 (blue) is measured with the system heated to $15$~K, which leads to an increase in conductance. Curve 3 (red) is a typical trace measured with the nanotube exposed to microwave radiation, while the substrate is kept at $5$~K. Two effects are clearly visible: an increase in slope under both conventional heating and applied microwave radiation, and, strikingly, a strong zero-current offset voltage appearing only with applied radiation. This zero-current offset voltage increases with increasing microwave power (upper scale in Fig.~2). We will argue below that these two effects observed with radiation are due to an increase of the electron temperature in the suspended carbon nanotube.

\begin{figure}[h!]
\includegraphics{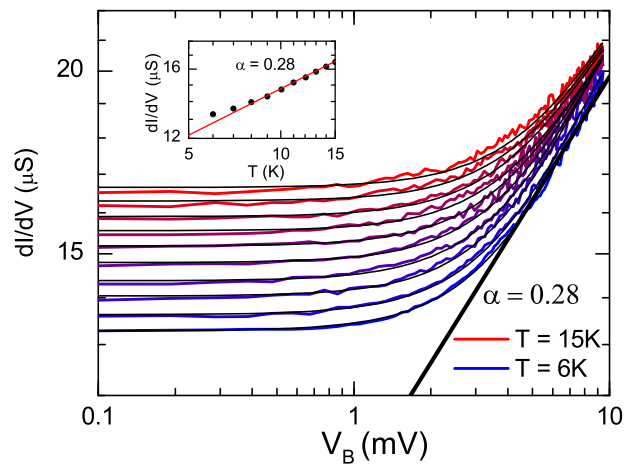}
\caption{\label{fig:T} (Color online) Power-law scaling of the differential conductance with bias voltage and temperature, indicative of a Luttinger liquid. The different curves correspond to an increase in the sample temperature from 6~K (bottom, blue) to 15~K (top, red). The thick black line shows the power-law scaling with bias voltage at high bias, with a scaling exponent of $\alpha = 0.28$. The black curves are fits from Eq.~(3). The inset shows the zero-bias differential conductance as a function of temperature, where the red line is a power law fit to the high temperature data, giving $\alpha = 0.28$.}
\end{figure}

First, we study the DC differential conductance, as a function of voltage and at different temperatures (Fig.~3), to determine the nature of the electronic states in the sample. The temperature is varied by a heating resistor on the copper block on which the sample is mounted. The colored curves correspond to increasing temperatures from bottom (blue) to top (red) with steps of 1 K between each curve. The conductance at low voltages increases with temperature. Similarly, the conductance increases with voltage beyond a voltage of about $\mathrm{1~mV}$. We interpret these data in the context of tunneling into a Luttinger liquid. In this theory, both dependencies will show scaling with a power law with identical exponents $\alpha$: at low voltages $(k_{\mathrm{B}}T \gg eV)$ we expect $\frac{dI}{dV} \propto T^{\alpha}$ and at high voltages $(eV \gg k_{\mathrm{B}}T)$ we expect $\frac{dI}{dV} \propto V^{\alpha} $, where $\frac{dI}{dV}$ is the differential conductance, $V$ the bias voltage, $T$ the temperature and $\alpha$ an exponent, which is determined by the strength of the interaction between the electrons in the one-dimensional wire.\cite{Matveev,KaneFisher} Applied to the data of Fig.~3, we find for $k_BT \gg eV$ power law scaling of the zero-bias differential conductance with temperature, with an exponent $\alpha = 0.28 \pm 0.05$ (red line in inset). The solid black line in the main figure shows power law scaling with bias voltage with the same exponent $\alpha = 0.28$.

As shown by Bockrath et al. \cite{Bockrath} for tunneling from a metallic lead at zero temperature to a Luttinger liquid at finite temperature $T_\mathrm{LL}$, the full set of differential conductance curves is given by
\begin{align}
f(V) &= \frac{dI}{dV}\bigg|_{T_{\mathrm{R}} = 0} \\
 &= AT_{\mathrm{LL}}^{\alpha}\cosh(\gamma\frac{eV}{2k_\mathrm{B}T_{\mathrm{LL}}})|\Gamma(\frac{1+\alpha}{2}+\gamma\frac{ieV}{2\pi k_{\mathrm{B}}{T_{\mathrm{LL}}}})|^2 \nonumber
\label{eq:1}
\end{align}
where $A$ is a constant, and $\Gamma$ is the complex gamma function. The parameter $\gamma$ is the ratio of the resistances of the two tunnel contacts of the lead-nanotube-lead system, a measure of an experimentally unavoidable asymmetry. In the limits of low and high bias Eq.~(1) reduces to the limiting values discussed above. In reality, the reservoirs also have a finite temperature $T_\mathrm{R}$, leading to a broadening in the differential conductance. Therefore, Eq.~(1) should be convoluted with the derivative of the Fermi distribution in the reservoirs, which is a bell-shaped curve of width $k_\mathrm{B}T_\mathrm{R}$, given by
\begin{equation}
g(V) = (1/4k_{\mathrm{B}}T_\mathrm{R})\mathrm{sech}^2(\gamma eV/k_{\mathrm{B}}T_\mathrm{R})
\label{eq:2}
\end{equation}
The experimentally measurable differential conductance is then given by:
\begin{equation}
\frac{dI}{dV}(V_\mathrm{B}) = \int_{-\infty}^\infty{f(V_\mathrm{B}-V)g(V)\mathrm{d}V}
\label{eq:3}
\end{equation}
where $V_\mathrm{B}$ is the applied bias voltage. In equilibrium, the temperature of the reservoirs and the temperature in the nanotube are identical. However, since we will argue below that our experimental results under microwave radiation can be understood by assuming differences in these temperatures, we explicitly use different temperatures in Eqs.~(1) and (2). In Fig.~3 the calculated differential conductance, for different temperatures of the entire substrate-lead-nanotube system, is shown (black curves), with the assumption that $T_\mathrm{LL}=T_\mathrm{R}$. The average asymmetry parameter $\gamma$ from these fits is $0.48 \pm 0.09$, and A is adjusted to match the data at $V_\mathrm{B} = 0$.

\begin{figure}[h!]
\includegraphics{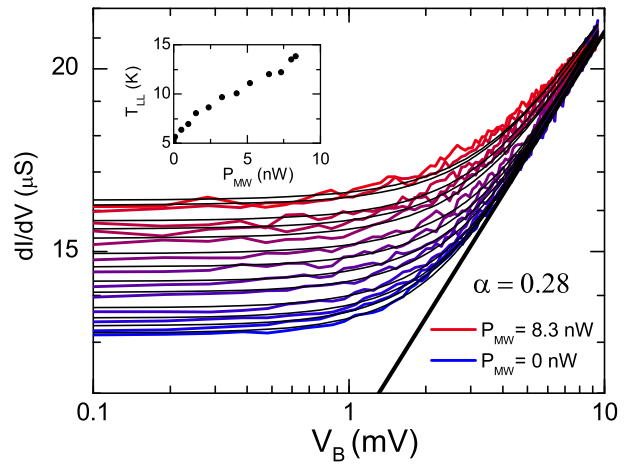}
\caption{\label{fig:LO} (Color online) Differential conductance of the suspended carbon nanotube as a function of bias voltage under increasing microwave power. The different curves correspond to an increase in the incident 108~GHz power from 0~nW (bottom, blue) to 8.3~nW (top, red). The black curves are fits from Eq.~(3), with $T_\mathrm{LL}\neq T_\mathrm{R}$ The inset shows the effective temperature determined from the zero-bias conductance, as a function of incident microwave power.}
\end{figure}

We now turn to the response to microwave radiation. Fig.~4 shows the differential conductance as a function of bias voltage for increasing incident power at 108 GHz. The colored curves correspond to increasing microwave power from bottom (blue) to top (red). The evolution of these conductance curves is analogous to that shown in Fig.~3, although the bath temperature is kept at a fixed value of 4.85~K. Given the macroscopic dimensions of the metallic contacts, we estimate that they will stay within $1~\mathrm{\mu K}$ of the bath temperature given the microwave power used. However, the thermal conductance out of the suspended nanotube is small, since energy can only flow out through the tunnel junctions at the ends. Therefore, we assume that the increase in conductance is due to an increase in electron temperature $T_{\mathrm{LL}}$ of the nanotube, and that the temperature $T_\mathrm{R}$ in the reservoirs stays at 4.85~K. With these assumptions the data are well described by Eq. (3). From the fits we find an average value of $\gamma = 0.46 \pm 0.06$ and $\alpha = 0.24 \pm 0.02$. The inset shows $T_{\mathrm{LL}}$ determined from the zero-bias conductance. It increases from 5~K to 14~K with increasing incident power. However, as opposed to the data in Fig.~3, we now have a non-equilibrium situation where the electron temperature in the nanotube is substantially higher than the temperature in the contacts.

The main temperature drop occurs over the tunnel barriers at the contacts of the nanotube and the leads, because of the relatively good thermal conductance along a nanotube and the high resistance of the tunnel barriers. In the nanotube growth process we obtain nanotubes with total resistances typically varying from 50~k$\Omega$ to 300~k$\Omega$. Because of the random nature of the growth process, these resistances will typically be distributed differently over the two different ends of the nanotube.

Having established that a consistent interpretation of the data is possible by assuming a different temperature for the electrons in the nanotube from those in the leads, we return to the offset voltage under microwave radiation, which was shown in Fig.~2. In principle, an offset-voltage can be generated by rectification of the microwave field by the non-linearity of the carbon nanotube current-voltage characteristic.\cite{SantaviccaArxiv,Rodriguez-Morales} However, given the measured nonlinearity, we estimate that an AC voltage of more than 12 mV is needed to obtain results on the order of the observed effect. This corresponds to an incoming microwave power of about 90 nW, given the microwave impedance of the system of 12.5 $k\Omega$ and irradiation from free space. This is an order of magnitude larger than the maximum incident microwave power of 8.3 nW. Instead, we argue that the offset-voltage arises from the strong temperature difference across the barriers between the nanotube and the contacts.

Under a temperature difference hot charge carriers diffuse from the nanotube to the colder reservoirs. This current will build up an electric field over the region where the temperature gradient is present, leading to an offset voltage at zero current. The voltage is assumed to be given by
\begin{equation}
V_{\mathrm{0}} = -S\Delta T \label{eq:Seebeck1}
\end{equation}
where $\Delta T$ is the temperature difference and $S$ is the thermopower (Seebeck-coefficient). A strong temperature gradient occurs primarily at the nanotube-contact barriers. In a perfectly symmetric sample, a thermovoltage would develop at both ends with opposite sign, and the net detected voltage would be zero. However, with an asymmetry in the contact resistances, the two thermovoltages will be different as well and a net voltage remains.

It has been predicted that $S$ is enhanced in a Luttinger-liquid system coupled to leads of non-interacting electrons:\cite{Krive}
\begin{equation}
S_{\mathrm{LL}}(T) = C_{\mathrm{S}}(g)S_{\mathrm{0}}(T) \label{eq:Seebeck2}
\end{equation}
where $S_{\mathrm{0}}(T)$ is the thermopower for non-interacting electrons. The thermopower is enhanced by a factor $C_{\mathrm{S}}(g) > 1$ for $g < 1$. Here $g$ is the Luttinger-liquid parameter measuring the strength of the interactions, which is directly related to the power law scaling exponent $\alpha$. For an end-contacted nanotube $\alpha = (g^{-1} -1)/4$ while for a bulk-contacted nanotube $\alpha = (g^{-1}+g-2)/8$.\cite{KaneFisher}

The increase in offset voltage as a function of microwave power, shown in Fig.~2, reflects the increase in temperature difference between the nanotube and the metallic leads. The lower axis in the inset gives the temperature difference determined from the differential-conductance measurements. In our experiments, the effective thermopower $S_{\mathrm{LL}}$ is constant over the temperature range studied. It is found to be $43~\pm 1~\mu\mathrm{V/K}$, which is an order of magnitude larger than the values reported for a carbon nanotube on a substrate.\cite{Small}

In conclusion, we have shown that we can selectively raise the temperature of the electron system in a single suspended carbon nanotube by microwave irradiation, without heating the electrodes. This results in both an increase of the differential conductance and the development of a strong thermovoltage due to the temperature gradients occurring locally at the contacts between the nanotube and the leads.

\textit{We acknowledge helpful discussions with B.H. Schneider, G.A. Steele, D.F. Santavicca, and D.E. Prober. We thank J. Barkhof of SRON for providing the microwave source. We like to acknowledge Microkelvin (No. 228464, Capacities Specific Programme) for financial support. This work is part of the research programme of the Foundation for Fundamental Research on Matter (FOM), which is part of the Netherlands Organisation for Scientific Research (NWO).}
\\

\end{document}